# The CALMA system: an artificial neural network method for detecting masses and microcalcifications in digitized mammograms

A. Lauria [a,?], R. Palmiero[a], G. Forni[a], P. Cerello[b], B. Golosio[c,d], F. Fauci[e], R. Magro[e], G. Raso[e], S. Tangaro[c,f], P.L. Indovina[a]

[a]*Università "Federico II" and Sezione INFN di Napoli, via Cinthia, 80126 Napoli, Italy*

[b]*Sezione INFN di Torino, via P. Giuria, 10125 Torino, Italy*

[c]*Sezione INFN di Cagliari, 09042 Monserrato (CA), Italy*

[d]*Università di Sassari, via Vienna, 07100 Sassari, Italy*

[e]*Università di Palermo, viale delle Scienze, 90128 Palermo, Italy*

[f]*Università di Bari, via Amendola, 70126 Bari, Italy*

**Abstract**

The CALMA (Computer Assisted Library for MAmmography) project is a five years plan developed in a physics research frame in collaboration between INFN (Istituto Nazionale di Fisica Nucleare) and many Italian hospitals. At present a large database of digitized mammographic images (more than 6000) was collected and a software based on neural network algorithms for the search of suspicious breast lesions was developed. Two tools are available: a microcalcification clusters hunter, based on supervised and unsupervised feedforward neural network, and a massive lesions searcher, based on a hibrid approach. Both the algorithms analyzed preprocessed digitized images by high frequency filters. Clinical tests were performed to evaluate sensitivity and specificity of the system, considering the system as alone and as secon reader. Results show that the system is ready to be implemented by medical industry. The CALMA project, just ended, has its natural development in the GPCALMA (Grid Platform for CALMA) project, where distributed users join common resources (images, tools, statistical analysis).

[?] Corresponding author: Tel.: ++39 081 676 121;  fax: ++39 081 676 346;  e-mail: adele.lauria@na.infn.it.

## 1. Introduction

In recent years different computerized systems have been developed to support diagnostic work of radiologists in mammography [1]. The goal of these systems is to focus the radiologist's attention on suspicious areas. They work in three steps: i. analogic mammograms are digitised; ii. images are segmented and preprocessed; iii. Regions of Interests (ROI) are found and classified by neural networks.

At present, many CAD systems are commercially available, but only two of them obtained the FDA approval. Preliminary studies showed that the use of computerized detection in mammography lead to a significant increment in sensitivity, both for masses and for microcalcification clusters [2, 3].

In the present work we report the performance of CALMA system in the search of cancer lesions in terms of sensitivity and specificity. Moreover it was tested as second reader in the support of the diagnostic work of the radiologists, evaluating how their sensitivity and specificity change with and without the CALMA system.

Results show that CALMA is ready to be used in the clinical practice as second reader. Moreover the large number of images collected in many hospitals suggests to realize a distribuited system for remote consultation of images. This is an application of the GRID technology that allows remote users to access common resources, like software, images, data.

## 2. Material and Methods

The CALMA hardware is composed by a personal computer (operative system: linux) and by a CCD linear scanner. Mammograms are first digitised (85?m, 12 bit/pixel), then are preprocessed by high frequency filters and saved in a special format.

Two different approaches are followed for the mass or the microcalcification cluster research. The microcalcification analysis is performed using two different Neural Networks. The first one is a FFNN that classifies windows in which images are segmented, the output of this NN becames the input for the second NN, that using the principal component method classifies ROIs. If the output value exceeds a threshold, the selected areas are pointed by a marker, as shown in figure 1.

The analysis of masses consists of three steps. In the first step not interesting data are eliminated; then an image analysis in the frequency domain is performed, at the end interesting areas are classified as ROIs by a FFNN.

Further software and hardware characteristics of the CALMA system can be found in the literature [4, 5].

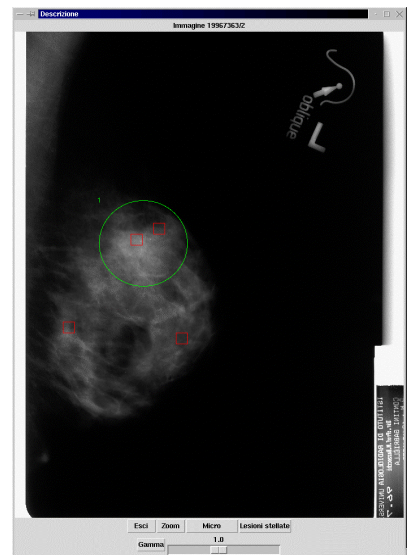

Figure 1. Output of the CALMA research of the lesions: the circle is the area containing the microcalcification cluster, and the squares are the ROIs pointed by the CALMA system.

A dataset composed by 180 images of healthy cases (with a three years follow up) and 145 images of malignant cases (with histopatological diagnosis) was considered to evaluate the sensitivity and the specificity of CAD in the search of masses. At the same way to test the system for microcalcification search 500 images of healthy cases (with a three years follow up) and 306 images of microcalcification clusters (with histopatological diagnosis) were collected. In each malignant case, the lesion was pointed by expert radiologists and the mean diameter is 2.1 cm for masses lesions and 2.3 cm for microcalcifications clusters. The system was evaluated in terms of sensitivity and specificity.

For a complete characterization of the CALMA system it is important to evaluate the diagnostic support to the radiologists. To this aim three radiologists (A, B, C) of different experience in mammography were considered (5, 3 and 2 years of experience in a public hospital, respectively). A dataset composed of 70 images of cancerous breasts (with microcalcification lesions) proven by biopsy and 120 images of healthy cases, proven by three years follow up was collected. Radiologists viewed all the images in the conventional way (looking films on diaphanoscope) indipendently each other and blind to the final diagnosis. After 8 months they reviewed images in a random order and with the help of CALMA system. The radiologists examine images on a screen using several visualization tools (e.g., zoom, contrast enhancement, etc.). When the microcalcification icon is clicked, the automatic search program starts. Programs containing both algorithms and neural networks perform the lesions analysis. In a few seconds, red markers point out regions of interest (ROI) where suspicious lesions are recognized by the system. To digitalize and visualize ROIs on images 2.5 minutes are necessary. To each ROI is associated a probability value, ranging from 0 to 1, of containing a lesion.

**Results**

In the research of masses the system get a sensitivity of 90% and a specificity of 85%; while for the microcalcification clusters 92% of sensitivity and 92% of specificity were obtained. In table 1 and 2 values of sensitivity and specificity, respectively, of radiologists without and with CAD were reported.

Meaningful results were obtained considering the increment in sensitivity of radiologists supported by CALMA, ranging from 10.0% (reader B) to 15.6% (reader C).

The results are comparable with other CAD systems commercially available, and show that CALMA system can be used as second reader and also that it can be industrially developed.

|   | Conventional | With CALMA |
|---|---|---|
| A | 82.8 % | 94.3 % |
| B | 80.0 % | 90.0 % |
| C | 71.5 % | 87.1 % |

Table1: sensitivity values of radiologists without and with CALMA System.

|   | Conventional | With CALMA |
|---|---|---|
| A | 87.5 % | 87.5 % |
| B | 91.7 % | 88.4 % |
| C | 74.2 % | 70.9 % |

Table2: specificity values of radiologists without and with CALMA System.